\documentclass[preprint,3p,times]{elsarticle}

\usepackage{amsfonts,amsmath,amsthm,amssymb}
\usepackage{graphicx,epsfig,pstricks,pst-node}
\usepackage{lineno}
\usepackage{figsize}
\usepackage{stfloats}
\usepackage{color,xcolor}

\journal{Chaos, Solitons \& Fractals}

\begin{document}

\begin{frontmatter}

\title{Shaping the learning signal in a combined Q-learning rule to improve structured cooperation}

\author[inst1]{Chunpeng Du}
\author[inst1]{Zongyang Li}
\author[inst2]{Yali Zhang}

\affiliation[inst1]{organization={School of Mathematics},
            addressline={Kunming University}, 
            city={Kunming},
            postcode={650214}, 
            state={Yunnan},
            country={China}}

\affiliation[inst2]{organization={School of Statistics and Mathematics},
            addressline={Yunnan University of Finance and Economics}, 
            city={Kunming},
            postcode={650221}, 
            state={Yunnan},
            country={China}}         
            
\author[inst2]{Yikang Lu}
\cortext[correspondingauthor]{Corresponding author.}
\ead{luyikang_top@163.com}

\affiliation[inst3]{organization={Institute of Technical Physics and Materials Science},
            addressline={Centre for Energy Research}, 
            city={Budapest},
            country={Hungary}}         

\author[inst3]{Attila Szolnoki\corref{correspondingauthor}}
\ead{szolnoki.attila@ek-cer.hu}
\begin{abstract}

Q-learning provides a standard reinforcement learning framework for studying cooperation by specifying how agents update action values from repeated local interactions outcomes. Although previous work has shown that reputation can promote cooperation in such systems, most models introduce reputation by modifying payoffs, encoding it directly in the state or changing partner selection, which makes it difficult to isolate the role of the learning signal itself. Here, we construct the reinforcement signal as a weighted combination of reputation and game payoffs, leaving the game and network structure unchanged. We find that increasing the weight on reputation generally promotes cooperation by consolidating clusters, but this effect is conditional on the learning dynamics. Specifically, this promoting effect vanishes in two regimes: when the learning rate is extremely small, which prevents effective information propagation and when the discount factor approaches one, as distant future expectations obscure the immediate reputational advantage. Outside these limiting cases, the efficacy of reputation in promoting cooperation is attenuated by higher learning rates but amplified by larger discount factors. These results advance the understanding of cooperative dynamics by demonstrating that cooperation can be stabilized through the reputational shaping of learning signals alone, providing critical insights into the interplay between social information and individual learning parameters.

\end{abstract}


\end{frontmatter}

\section{\label{sec:intro}Introduction}
Cooperation is a founding pillar of biological and social systems, supporting collective defense in animal groups, division of labor in eusocial species, public goods and in human societies \cite{nowak_11,hua_sj_prr24,shu_lm_prsa22,szolnoki2015antisocial}. Nevertheless, it is difficult for cooperation to emerge and persist because individuals can easily benefit from common goods without contributing to it. More precisely, contributors incur the cost, while the resulting benefits are shared independently of the contribution. As a response, the fear of potential exploitation discourages cooperators, which allows opportunistic behavior to spread \cite{bhui2019exploitation,han_jrsif22,huang_cc_csf24}. These difficulties have motivated sustained studies of how cooperation emerges and persists. Evolutionary game theory provides a standard framework for studying the emergence and stability of cooperation, linking individual incentives, interaction patterns, and population-level dynamics. \cite{sigmund1999evolutionary,wang2015evolutionary,wang_cq_amc24,liu_yy_jrsif25}. It specifies a payoff structure and a behavioral update rule at an individual level, allowing researchers to analyze how interactions governed by that rule produce collective population level dynamics, including steady state cooperation levels and characteristic spatial patterns \cite{perc_pr17,feng_m_csf23,ohtsuki2006simple,floria_pre09}.

A substantial body of evolutionary game theory research has identified mechanisms that promote or sustain cooperation under specified conditions \cite{he_s_csf26,pan_amc24}. Nowak summarized five classical cases: kin selection, direct reciprocity, indirect reciprocity, network reciprocity and group selection \cite{nowak2006five,wang2013interdependent}. These mechanisms are widely used as benchmark explanations for the evolution of cooperation in evolutionary game theory. Among these, network reciprocity is fundamental when interaction occurs in fixed structures \cite{szolnoki2014facilitators,su_r_csf24}. In particular, local interaction allows cooperators to form homogeneous clusters that reduce exposure to exploitation and stabilize cooperative domains on networks \cite{bin_l_amc23,kang_hw_pla24}. Analytical work on regular networks has established a simple selection condition, expressed as the benefit to cost ratio and the degree, which predicts when cooperation spreads under local updating \cite{wang2024evolutionary,fotouhi2019evolution}. With the development of complex network science, the dynamics of cooperative behavior has been examined in more realistic structures, including heterogeneous networks and interdependent networks \cite{yan_zy_csf24,boccaletti2006complex}. In addition to structural effects, mechanisms such as rewards, punishments, memory and expectations have been shown to enhance group cooperation in networks by promoting network reciprocity \cite{yan_ry_pla24,geng2025evolution,guo2024evolution}.

In structured networks, interaction is local and repeated within small neighborhoods, making a record of past behavior especially consequential \cite{sampson2002assessing,perc2013evolutionary}. Reputation serves as key information at individual level that acts as a social signal summarizing recent behavior. Depending on how this information is shared and the applied update rule, it can diffuse locally and influence decisions at cooperator defector interfaces, thereby affecting the stability of cooperative domains \cite{fu2008reputation,sun_xp_pla25,tian2016cooperation,ding2016reputation}. In networks, a body of work instantiates reputation and measures its cooperative impact. In co-evolving social networks, reputation based partner choice sustains cooperation by redirecting ties toward well rated neighbors \cite{fu2008reputation,szolnoki2009resolving,shi2024enhancing}. Locally available image scoring rules increase cooperation when neighbors can observe or infer the reputation state \cite{tian2016cooperation,nax_srep15b,brandt2005indirect}. In spatial models with second-order reputation, evaluation rules that depend on both a player’s action and the recipient’s reputation result in a higher cooperation level than spatial reciprocity alone \cite{wang_p_csf24,dong2019cooperation,zhao_cy_c24}. Laboratory experiments further show that gossip mediated diffusion of reputation information and the possibility of ostracism increase cooperation in group settings \cite{feinberg2014gossip,xu_ws_csf26}.

The aim of our work is to integrate reputation information into value based learning on networks. Some studies encoded the reputation directly in the reward, so agents update action values from returns shaped by their reputation level. These models typically find greater cooperation when the assessment rules are consistent and locally observable \cite{yao2024digital,zhang2024impact,ren2023reputation}. Other works encoded the reputation in the agent’s state, so the Q-values are updated conditional on the reputation level and reported that richer assessment rules, such as second-order reputation, further enhance cooperation on lattice structures \cite{xie_k_csf25,dong2019cooperation}. There is also evidence that tailoring the learning signal to local interaction improves cooperation relative to imitation updates on networks \cite{szolnoki2020gradual,zare2024survey}. Laboratory and multi-agent studies show that reputation cues, when observable and evaluated according to stable rules, are used by value-based learners to adjust their strategies and support cooperation \cite{feinberg2014gossip,quan_j_jsm22}.

Most of the existing combinations of the reputation and the Q-learning protocol change the payoffs in the stage game or modify partnership interactions when introducing reputation \cite{xie_k_amc26,yue2025dynamic}. Other works encode the reputation directly in the agent’s state using custom features, a method that adds additional parameters and obscures which component produces the gains \cite{pinyol2013computational}. In both cases, it becomes challenging to isolate the effect of learning from the reputation information and the relative influence of the game payoff and reputation is often not on a common numerical scale \cite{wei_x_epjb21,dellarocas2006reputation}. In contrast, we define the reinforcement signal as a weighted combination of reputation and game payoffs, rather than using reputation to alter either the strategic payoffs or interaction partners. This design follows the standard reward shaping practice by guiding learning without changing the underlying game and keeping the role of reputation explicit \cite{ibrahim2024comprehensive,xiao_j_jpc23}.

In our setting, reputation and game payoff are combined in a single reinforcement signal that updates Q-values in the spatial prisoner’s dilemma. This setup gives cooperator clusters a local advantage at their boundaries with defectors and, in turn, strengthens network reciprocity. As a result, cooperation rises monotonically with the weight placed on reputation. Two limits qualify this trend. When the learning rate is very small, the reputation advantage propagates too slowly along boundaries and the gain vanishes. When the discount factor is very large, returns far in the future are weighted almost as strongly as immediate ones, which diminishes the local advantage generated by reputation and again removes the cooperative gain. Outside these limiting cases, two regularities hold. The promoting effect of reputation decreases as the learning rate increases, because higher learning rates place more weight on the most recent return and diminish the influence of accumulated reputation. Conversely, the promoting effect of reputation strengthens as the discount factor increases, because a larger discount factor allows reputation advantages to persist across updates, translating them into enhanced network reciprocity.

\section{\label{sec:model}Model}
We consider an $L \times L$ square lattice with periodic boundary conditions, where each node represents an agent that interacts with its four nearest neighbors (von~Neumann neighborhood $\Omega_i$ of size $k=4$). In each round, every agent plays a prisoner’s dilemma game with its four neighbors and accumulates the total payoff from all pairwise interactions.
We use the weak PD parametrization \cite{rapoport1988experiments,nowak_n92b}:
\begin{equation}
A = \begin{pmatrix} 1 & 0 \\ b & 0 \end{pmatrix}, \quad 1<b<2,
\label{eq1}
\end{equation}
where the payoff ranking satisfies $T=b>R=1>P=S=0$ and $2R>T+P$. Pure strategies are represented as column vectors:
\begin{equation}
S_C = \begin{pmatrix} 1 \\ 0 \end{pmatrix} \text{ or } S_D = \begin{pmatrix} 0 \\ 1 \end{pmatrix}.
\label{eq:Si}
\end{equation}

The direct game payoff of agent $i$ at round $t$ is given by:
\begin{equation}
F_i = \sum_{j \in \Omega_i} S_i A S_j^T,
\label{eq3}
\end{equation}
which represents the sum of the pairwise payoffs obtained from interactions with all neighbors. Each agent $i$ maintains a reputation variable $r_i \in [0,r_{max}]$ with $r_{max}=100$. Reputation evolves deterministically according to the agent’s action:
\begin{equation}
r_i(t+1) = 
\begin{cases} 
r_i(t) + 1, & \text{if } S_i(t) = C, \\
r_i(t) - 1, & \text{if } S_i(t) = D, 
\end{cases}
\label{eq4}
\end{equation}
if \( r_i(t+1) > 100 \), the value is capped at \( 100 \), while if \( r_i(t+1) < 0 \), it is reset to \( 0 \). We define the normalized reputation score as:
\begin{equation}
R_i(t) = \frac{r_i(t)}{r_{\text{max}}}.
\label{eq5}
\end{equation}

Because 
$F_{i}(t) \in [0,4b]$, 
we also normalize the direct payoff:
\begin{equation}
\pi_{i}(t)= \frac{F_i(t)}{4b}.
\label{eq6}
\end{equation}

The total reward is defined as a weighted combination of payoff and reputation:
\begin{equation}
\Pi_{i}(t) = (1 - \beta) \times \pi_{i}(t) + \beta \times R_i(t),
\label{eq7}
\end{equation}
where the reputation weight \(\beta \in [0,1]\) controls the relative influence of social information on the final reinforcement signal.

Each agent acts as a reinforcement learner with state space $S=\{C,D\}$ and action space $A=\{C,D\}$. The state encodes the agent's own previous action, hence if the agent takes action $a_t \in A$ in round $t$ then the next state is $s_{t+1} = a_t$. The Q-values, denoted $Q^i_{(s,a)}(t)$ are updated through standard Q-learning and Q-table for agent $i$ is:
\begin{equation}
Q^i_{(s,a)}(t) = \begin{pmatrix}
Q^i_{(C,C)}(t) & Q^i_{(C,D)}(t) \\
Q^i_{(D,C)}(t) & Q^i_{(D,D)}(t)
\end{pmatrix},
\label{eq8}
\end{equation}
where \( Q^i_{(s,a)}(t) \) denotes the expected return for selecting action \( A \) in state \( S \) in round \( t \). The Q-values are iteratively updated according to the rule:
\begin{equation}
Q^i_{(s,a)}(t+1) = (1 - \alpha)Q^i_{(s,a)}(t) + \alpha \left[ \Pi_{i}(t) + \gamma \max Q^i_{(s',a')}(t) \right],
\label{eq9}
\end{equation}
where learning rate $\alpha \in(0,1)$ and discount factor \( \gamma \in (0,1) \). To avoid premature convergence to suboptimal policies, we employ an \( \epsilon \)-greedy exploration strategy, where an agent chooses the action with the highest Q-value with probability \( 1 - \epsilon \) or selects a random action with probability \(\epsilon \).

In this work, we use synchronous strategy updates. At the beginning, each agent is randomly assigned to cooperate or defect with equal probability and receives a random initial reputation within [0, 100]. All Q-values are initialized to $0$. Each Monte Carlo step consists of four sequential stages. First, each agent selects an action from its Q-table using the \( \epsilon \)-greedy rule. Second, it plays the prisoner’s dilemma with its four neighbors and accumulates the resulting payoff values. Third, reputation is updated and to form the reinforcement signal. Fourth, it updates Q-values and sets the next state. The simulation proceeds for 100,000 Monte Carlo steps and the cooperation level $\rho_C$ is computed as the average over the final 5,000 steps.

\section{\label{sec:result}Results}

\begin{figure}[htbp]
    \centering
    \includegraphics[width=0.4\linewidth]{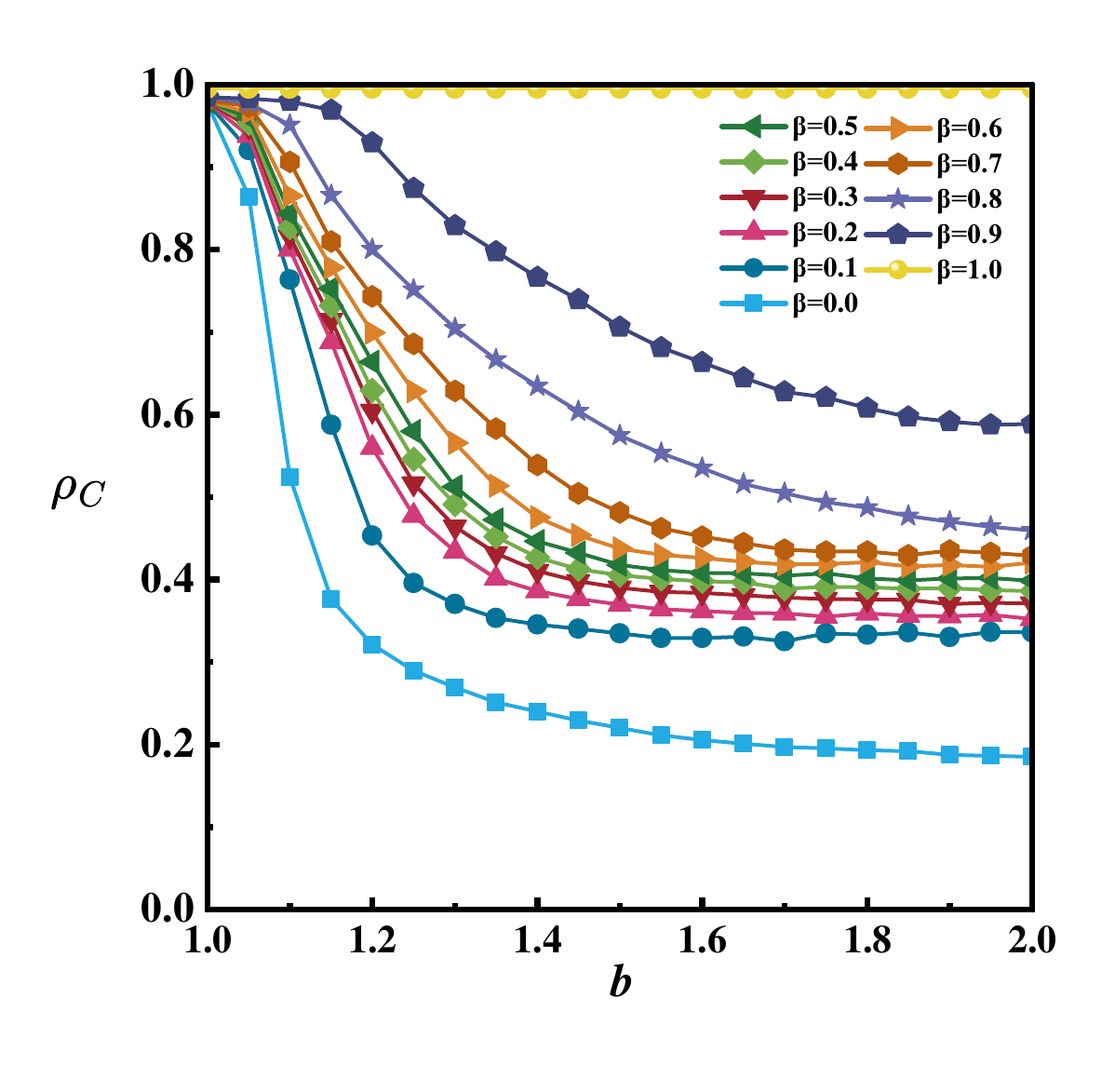}
    \caption{The frequency of cooperators $\rho_c$ in dependence of the temptation to defect $b$, under various values of the reputation weight $\beta$. The values of $\beta$ range from [0,1] as indicated in the legend. All results are obtained under the parameter settings of $\alpha = 0.1$, $\gamma = 0.95$ and $\epsilon = 0.01$. The figure suggests that a larger reputation weight has a positive consequence on general cooperation.}
    \label{fig:b-dependence}
\end{figure}

We first examine how the reputation weight $\beta$ influences the cooperation level under different dilemma strengths $b$. The results are summarized in Fig.~\ref{fig:b-dependence}. We find that regardless of the value of $b$, the greater the reputation weight $\beta$, the higher the fraction of cooperation. Even when $b$ is relatively large, cooperative behavior persists under the influence of $\beta$. This phenomenon suggests that reputation plays a stronger role than game payoffs in maintaining cooperative behavior when reinforcement signal is used. As the reputation weight in the reinforcement signal increases, cooperative behavior is more likely to be maintained and developed.

\begin{figure}[h]
    \centering
    \includegraphics[width=0.7\linewidth]{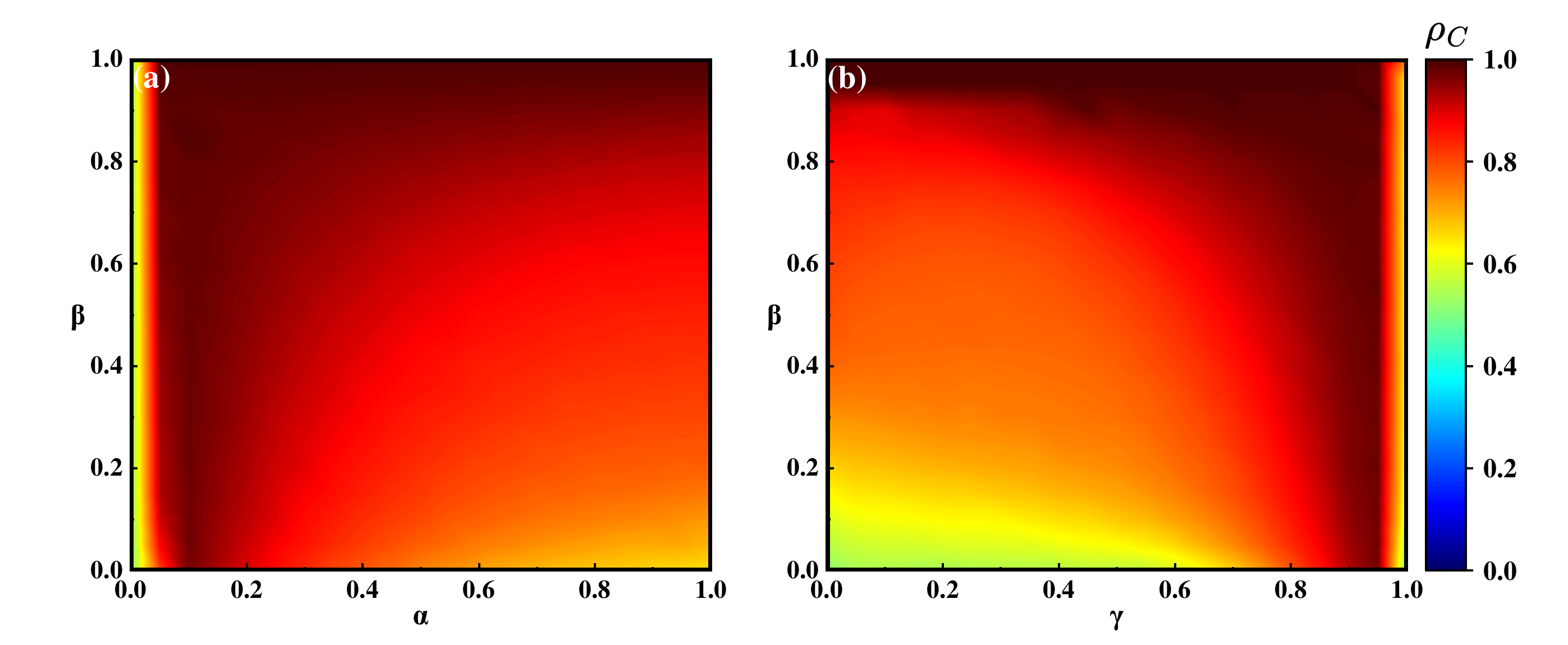}
    \caption{The cooperation level on the parameter plane of $\alpha$ and reputation weight (a) and on the parameter plane of discount factor \(\gamma\) and reputation weight (b). The color-coded stationary values of $\rho_C$ are indicated by the bar shown on the right-hand side. While the effect of parameter $\beta$ on $\rho_C$ is straightforward, it cannot be said about the effects of other parameters $\alpha$ and $\gamma$. We can identify an intermediate optimal $\alpha$ and $\gamma$ values in both panels where the highest cooperation level can be reached. Other parameters are set as \(b = 1.02\) and $\epsilon = 0.01$.}
    \label{fig:alph-bet}
\end{figure}

To get a more complete overview about the possible role of learning parameters, Fig.~\ref{fig:alph-bet} shows the cooperation level in the $(\beta, \alpha)$ and $(\beta, \gamma)$ planes. In both planes, cooperation generally increases with $\beta$, indicating that a stronger reputation influence in the reinforcement signal expands the region of high cooperation. In the $(\beta, \alpha)$ plane (Fig.~\ref{fig:alph-bet}(a)), the reputation effect vanishes when the learning rate is extremely small ($\alpha \to 0$): Q-values change very slowly and cooperation remains low for all $\beta$ values. For $\alpha$ above this boundary, increasing $\beta$ raises the global cooperation level. However, the impact of reputation gradually diminishes as $\alpha$ increases, with the highest cooperation observed at large $\beta$ and intermediate $\alpha$. In the $(\beta, \gamma)$ plane (Fig.~\ref{fig:alph-bet}(b)), cooperation shows a non-monotonic dependence on the discount factor $\gamma$. For intermediate $\gamma$, increasing $\beta$ again expands the high cooperation region, whereas for $\gamma$ very close to 1 the reputation weighted reinforcement signal no longer sustains cooperation, and the system returns to low cooperation across the plane. Overall, Fig.~\ref{fig:alph-bet} identifies the ranges of the learning parameters within which reputation effectively promotes cooperation in the system. 
\begin{figure}[h]
    \centering
    \includegraphics[width=0.8\linewidth]{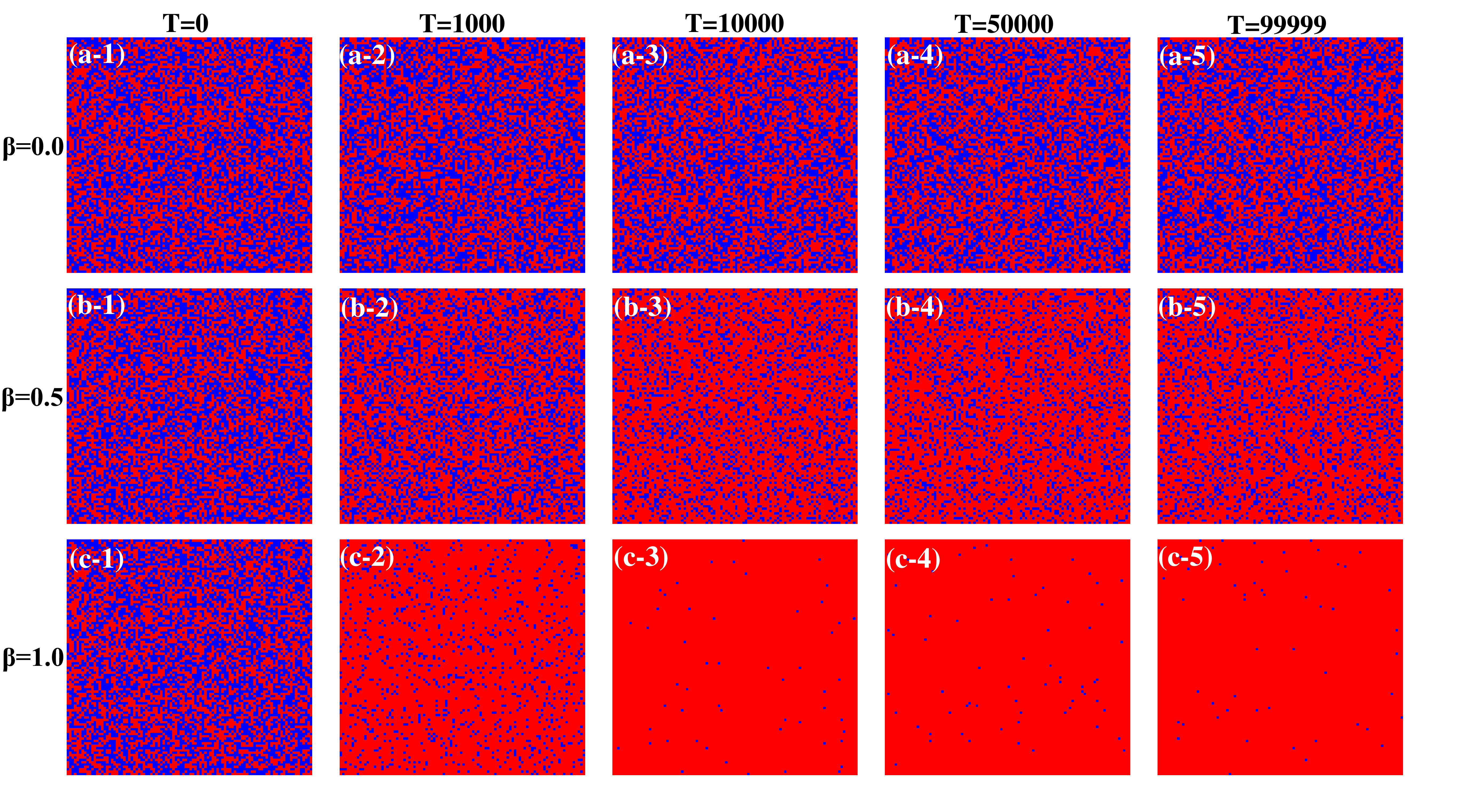}
    \caption{The time evolution of spatial patterns at different reputation weight \(\beta\). From top to bottom, the values of \(\beta\) are 0.0, 0.5 and 1.0. The snapshots were taken at time steps \(T = 0, 1000, 10000, 50000\) and \(99999\). Defectors and cooperators are represented by blue and red cells, respectively. As expected, larger $\beta$ value provides an increased cooperator community. These players, however, do not form as compact clusters as we could previously observed when imitation dynamics was used. The remaining parameters are set as \(\alpha = 0.8\), \(\gamma = 0.6\), \(b = 1.05\) and $\epsilon = 0.01$.}
    \label{fig:snapshots-bet}
\end{figure}

To understand the system behavior more deeply, we present the spatiotemporal evolution patterns in Fig.~\ref{fig:snapshots-bet} to illustrate the microscopic mechanism through which reputation promotes cooperation. Without reputation ($\beta = 0$), the lattice remains mottled and the domain walls are unstable, consistent with the low cooperation observed in Fig.~\ref{fig:b-dependence}. At $\beta = 0.5$, cooperators gradually begin to form larger clusters in time. However, small, compact defector clusters still persist by exploiting cooperators at the boundaries. In other words, the cooperator domains are not as homogeneous as previously observed when the strategy update was driven by the imitation rule based on pairwise comparison~\cite{szolnoki_pre11,szolnoki_pre11b}. Therefore, the less regular domain walls separating competing strategies can be considered as a characteristic benchmark of pattern formation driven by reinforcement learning protocol.  
When $\beta = 1.0$, cooperator clusters expand effectively. In this case, defectors have almost no ability to invade cooperators, can only survive in rarely scattered spots in the cooperator matrix, and the system maintains a highly cooperative state. These spatiotemporal evolution patterns indicate that increasing the reputation weight accelerates the consolidation and spread of cooperative clusters, enhances network reciprocity and thereby improves the fraction of cooperation.
\begin{figure}
    \centering
    \includegraphics[width=0.8\linewidth]{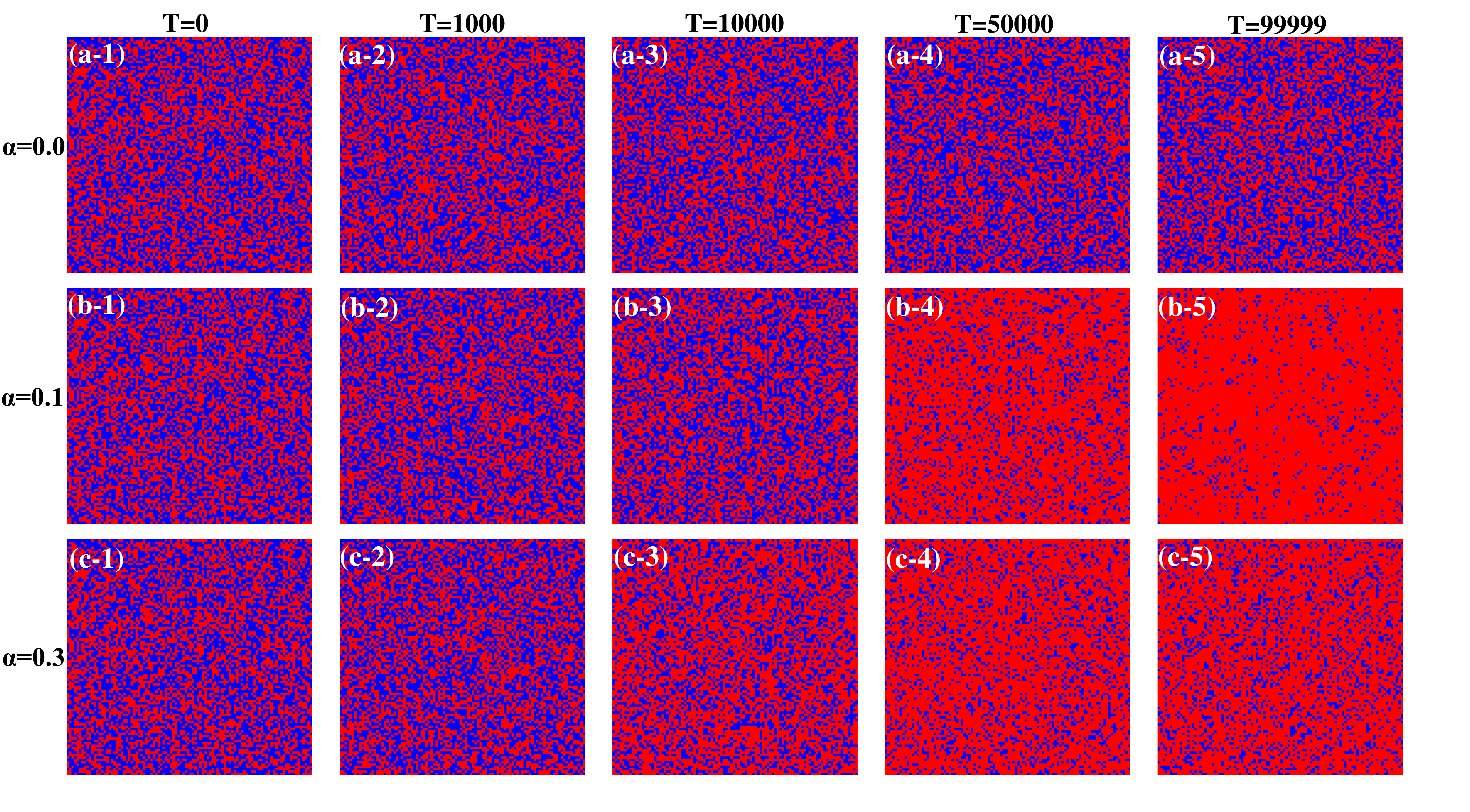}
    \caption{The time evolution of spatial patterns at different reputation weight $\alpha$. From top to bottom, the values of \(\alpha\) are 0.0, 0.1 and 0.3. The snapshots were taken at time steps \(T = 0, 1000, 10000, 50000\) and \(99999\). Defectors and cooperators are represented by blue and red cells, respectively. The non-monotonous consequence of increasing parameter $\alpha$ nicely demonstrated by the middle row where red cooperators dominate the system. The remaining parameters are set as \(\beta = 0.3\), \(\gamma = 0.95\), \(b = 1.05\) and $\epsilon = 0.01$.}
    \label{fig:snapshots-alp}
\end{figure}

For comparison, we also examine how the learning rate $\alpha$ shapes the effect of the reputation weight on cooperation. This can be easily done by presenting the related spatiotemporal patterns at some representative parameter values. In Fig.~\ref{fig:snapshots-alp}, we vary $\alpha$ while keeping $\beta$, $\gamma$ and $b$ fixed. At \(\alpha=0\), the system remains mixed and cooperator clusters do not expand over time. At \(\alpha=0.1\), cooperator clusters starts expanding. In the stationary state we practically have a huge cooperative matrix which contains small defector clusters. Similarly to our previous observations, the separating domain walls is not as sharp as for systems using imitation update, which is a clear consequence of the reinforcement learning rule applied here. At \(\alpha=0.3\), consolidation begins later and small defector clusters persist longer before disappearing. These patterns indicate that a nonzero learning rate is required for the reinforcement signal that blends normalized payoff and reputation to affect behavior. In the same time, a larger \(\alpha\) places more weight on the most recent return and slows the consolidation of cooperation. 
\begin{figure}
    \centering
    \includegraphics[width=0.8\linewidth]{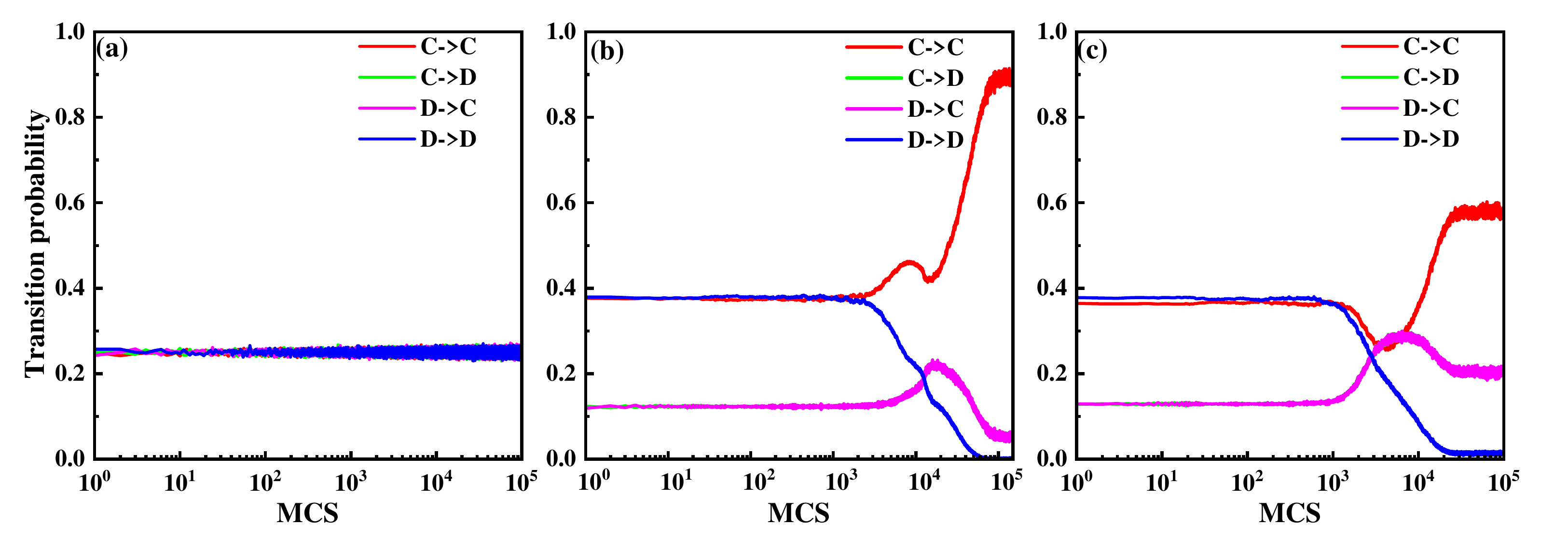}
    \caption{Time evolution of microscopic transition probabilities at different values of $\alpha$. The three panels show the cases obtained at $\alpha=0$, 0.1 and 0.3. As shown in the legend, red line represents the likelihood that a cooperator players remains cooperator, and blue line the probability when a defector keeps its strategy. Pink line depicts the probability that a defector becomes cooperator, while green line marks the opposite process. Note that green is hardly seen, covered by pink, because strategy change processes are equally frequent in the stationary state. Interestingly, the less intensive individual movements can be observed in the middle panel, which results in the highest general cooperation level. All data are obtained under the parameter conditions of \(\beta=0.3\), \(\gamma=0.95\), \(b=1.05\) and $\epsilon = 0.01$.}
    \label{fig:time}
\end{figure}

To support our argument more directly, we monitor how the transitions between competing strategies change in time. In particular, Fig.~\ref{fig:time} shows the four elementary strategy transitions over time at the three representative learning rates used in Fig.~\ref{fig:snapshots-alp}. When \(\alpha=0\), all transition probabilities remain close to their initial levels and the system shows no systematic shift. At this parameter value the Q-values do not change, therefore the reinforcement signal that blends payoff and reputation is never incorporated into behavior. As a consequence, reputation cannot accumulate into an advantage for cooperators, cooperator clusters do not consolidate and the population settles at a low cooperation level. At \(\alpha=0.1\), the probability of \(C \to C\) rises, \(C \to D\) and \(D \to C\) fall, which aligns with the formation and thickening of cooperator clusters. At \(\alpha=0.3\), the probability of \(C \to C\) is noticeably lower than at \(\alpha=0.1\) and \(C \to D\) is higher, indicating that cooperators are less able to maintain their status. This explains the decrease in cooperation observed in Fig.~\ref{fig:alph-bet}(a) at higher learning rates, where a larger $\alpha$ makes cooperator clusters less stable. In this case, the larger learning rate places more weight on the most recent return and less on accumulated reputation, therefore the reputation becomes less effective at promoting cooperation.
\begin{figure}[h]
    \centering
    \includegraphics[width=0.8\linewidth]{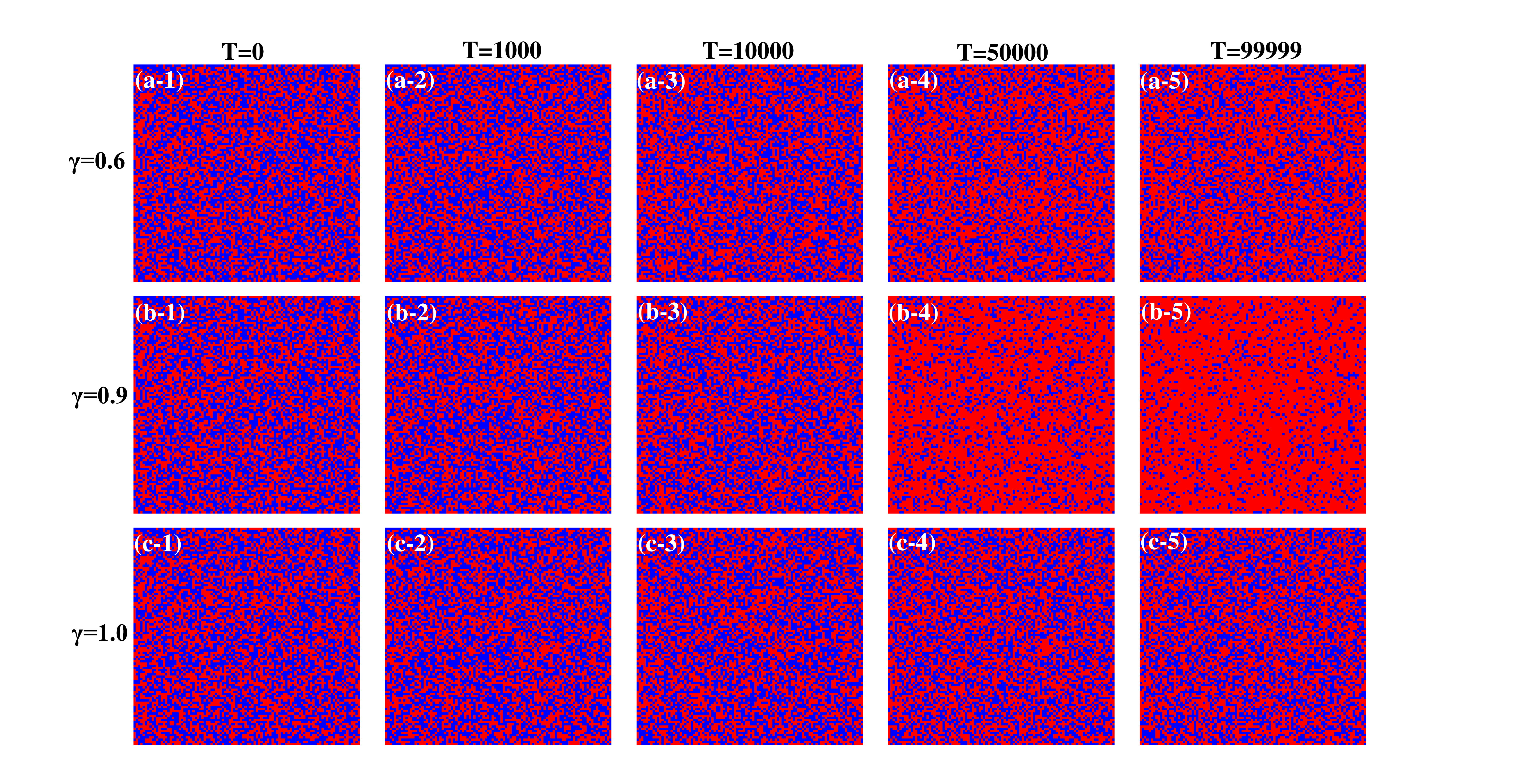}
    \caption{The time evolution of spatial patterns at different reputation weight $\gamma$. From top to bottom, the values of \(\gamma\) are 0.6, 0.9 and 1.0. The snapshots were taken at time steps \(t = 0, 1000, 10000, 50000\) and \(99999\). Defectors and cooperators are represented by blue and red cells, respectively. Tho optimal intermediate value of $\gamma$ can be easily noticed in the middle row. The remaining parameters are set as \(\beta = 0.6\), \(\alpha = 0.1\), \(b = 1.05\) and $\epsilon = 0.01$.}
    \label{fig:snapshots-gam}
\end{figure}

To make our study complete, we also present the representative spatiotemporal patterns obtained at different values of $\gamma$ parameter. In Fig.~\ref{fig:snapshots-gam} we show how the discount factor $\gamma$ shapes the impact of reputation on cooperation. In particular, we vary $\gamma$ while keeping $\beta = 0.3$, $\alpha=0.1$ and $b=1.05$ fixed. At $\gamma = 0.6$, although cooperators manage to form clusters as the system evolves, these clusters remain small and are fragmented by defector clusters. At \(\gamma=0.9\), cooperator clusters grow larger and become more connected, while defector clusters shrink and retreat into smaller patches. At \(\gamma=1\), cooperator clusters cannot grow and defector enjoy their vicinity at their boundaries, indicating a partial weakening the network reciprocity compared to the case obtained at $\gamma=0.9$. These snapshots indicate that cooperation is strongest at an intermediate value of \(\gamma\). With a moderate discount factor, the reputation signal remains relevant for several rounds and helps cooperator clusters grow. When the discount factor approaches one, agents weight future rewards too heavily, causing distant future expectations to obscure the immediate reputation advantage near the edges of clusters and cooperator clusters stop growing. 

\begin{figure}
    \centering
    \includegraphics[width=0.9\linewidth]{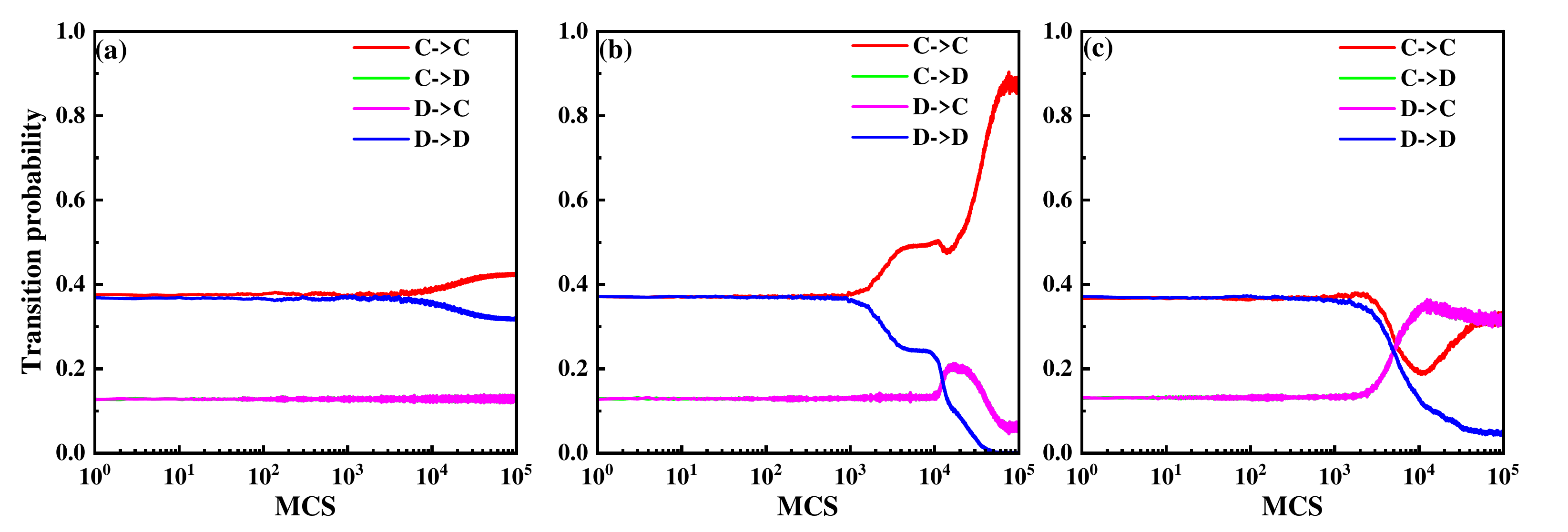}
    \caption{Time evolution of microscopic transition probabilities at different values of $\gamma$. The three panels show the cases obtained at $\gamma=0.6$, 0.9 and 1.0. The color codes are identical to those used in Fig.~\ref{fig:time}. As previously, the highest cooperation level is accompanied with the smallest exchange intensity between the competing strategies. All data are obtained under the parameter conditions of \(\beta=0.6\), \(\alpha=0.1\), \(b=1.05\) and $\epsilon = 0.01$.}
    \label{fig:time-trans}
\end{figure}

In order to support our observations from a different perspective we last present the time evolutions of the intensity of the four elementary strategy transitions \(C\to C\), \(C\to D\), \(D\to C\) and  \(D\to D\) at different values of $\gamma$. For a proper comparison we use similar parameter setting with $\alpha=0.1$, $\beta=0.6$ and $b=1.05$, as used for Fig.~\ref{fig:snapshots-gam}. Our results are summarized in Fig.~\ref{fig:time-trans}. At $\gamma =0.6$, all four transition probabilities remain close to initial state. At \(\gamma=0.9\), the probability of \(C\to C\) rises, the probability of \(D\to C\) becomes higher than that of \(C\to D\), \(D\to D\) declines, and \(C\to D\) remains relatively low. At \(\gamma=1\), the probability of \(C\to C\) declines and the probabilities of \(C\to D\) and \(D\to D\) increase compared with $\gamma=0.9$, indicating that cooperators are less able to maintain their strategy and defection becomes more persistent. It is worth noting that the smallest intensity of strategy exchange can be seen at the optimal value of $\gamma$ which provides the highest general cooperation for the system. This observation is in good agreement with the one we reported when the evolution of transition probabilities were shown in dependence of the alternative parameter $\alpha$.
Taken together, Figs.~\ref{fig:snapshots-gam} and \ref{fig:time-trans} support the non-monotonic dependence on $\gamma$ observed in Fig.~\ref{fig:alph-bet}(b). Highlighting that reputation promotes cooperation most effectively at an intermediate value of discount factor, while very large $\gamma$ weakens the influence of the reputation weight in the reinforcement signal and reduces its ability to sustain high cooperation.

\section{\label{sec:conc}Conclusion}

In this work, we explored how reputation affects cooperation in a spatially structured population when it is integrated into the reinforcement signal of Q-learning. Agents, distributed on a square lattice, update their actions using Q-learning based on a reward signal that combines normalized game payoffs with a simple bounded reputation score, weighted by a reputation parameter $\beta$, while the payoff matrix and the interaction network remain unchanged. In close agreement with the general expectation that social information promotes prosociality, we find that cooperation rises monotonically with the weight placed on the reputation signal. However, we find that this promoting effect is conditional and vanishes in two regimes. These parameter areas are identified when the learning rate is small, which prevents effective propagation of information between neighbors and when the discount factor approaches one, as distant future expectations obscure the immediate reputation advantage. Notably, outside these limiting cases, the efficacy of reputation in promoting cooperation is attenuated by higher learning rates but amplified by large discount factors. Overall, our findings deepen the understanding of how reputation-based information shapes network reciprocity and underscore the importance of accounting for the learning dynamics themselves, not just the presence of social information, when designing interventions.

Compared with previous studies that introduced reputation by changing payoffs or partner selection, our work focuses on a different aspect: shaping the learning signal itself. Prior applications of reinforcement learning in networked games have primarily focused on mechanisms that alter the strategic environment through reputation, such as reputation-based partner choice \cite{fu2008reputation,xu2024reinforcement,bi_p_ieee25}, payoff modifications (punishment/reward) \cite{szolnoki_epl10}, or co-evolutionary rules. By contrast, our model keeps the game and network fixed and only adjusts how agents combine game payoff and reputation inside the Q-learning rule. The results show that this internal weighting is sufficient to consolidate cooperative clusters on a lattice structure and to raise the cooperation level, provided that the learning parameters stay in an effective range. In this sense, the reputation weight in the reinforcement signal is not just a technical parameter but a meaningful control parameter for how strongly social information supports network reciprocity under reinforcement learning.

Beside the novel and instructive insights offered by our present work, several limitations remained for future research. First, in our present work we wanted to focus on the simplest model to identify the clear consequence of the signal we introduced. Therefore, we have chosen the simplest spatial structure of a lattice which characterizes a homogeneous population where all actors have equal neighbors. Future studies should extend this to heterogeneous networks \cite{wang2015evolutionary}, or interdependent networks to understand the impact of reputation-integrated Q-learning on cooperative behavior under complex topologies. Second, we employed a fixed reputation rule. Future work could explore more complex assessments, such as second-order reputation~\cite{guo_csf23} or partially observed reputation information. Finally, the agents in our model use non-evolving learning parameters. Future work could incorporate adaptive learning rates, which can simulate more realistic responses to reputation information. Addressing these factors is essential for assessing the robustness of the impact of reputation-based learning on cooperative behavior in real world systems.

\section*{Acknowledgments}

C.P. is funded by the Yunnan Fundamental Research Projects(No.202401AU070018), the Scientific Research Fund of the Yunnan Provincial Department of Education(No.2024J0774). Y.K. acknowledges support from the Yunnan Fundamental Research Projects(Grant Nos.202501AU070193), the Scientific Research Fund of the Yunnan Provincial Department of Education(Grant No.2025J0579) and the Foundation of Yunnan Key Laboratory of Service Computing(Grant No.YNSC24125). A. S. is supported by the National Research, Development and Innovation Office (NKFIH), Hungary under Grant No. K142948.

\bibliographystyle{elsarticle-num} 
\bibliography{references}

\end{document}